\pdfoutput=1
\documentclass[acmlarge]{acmart}
\usepackage{balance}  

\usepackage{bbm}  

\usepackage{amsmath}          
\usepackage{amssymb} 			    
\usepackage{textcomp}

\usepackage{url}              

\usepackage{array}            
\usepackage{tabularx}         
\usepackage{colortbl}
\newcolumntype{Y}{>{\centering\arraybackslash}X}	

\usepackage{mathtools}          

\usepackage{pbox}   

\usepackage{setspace}

\usepackage{algorithm}
\usepackage[noend]{algpseudocode} 
\usepackage{algorithmicx}

\algnewcommand{\COMMENT}[2][.5\linewidth]{\leavevmode\hfill\makebox[#1][l]{//~#2}}
\algnewcommand{\COMMENTT}[2][.5\linewidth]{\leavevmode\hfill\makebox[#1][l]{\hphantom{aa} // ~#2}}
\algnewcommand{\LineComment}[1]{\State \(//\) #1}	
\algnewcommand\RETURN{\State \textbf{return} }

\usepackage{graphicx}
\usepackage[space]{grffile}   
\graphicspath{{figs/}}
\DeclareGraphicsExtensions{.pdf,.jpeg,.jpg,.png}

\usepackage{caption}
\usepackage{enumitem}
\newcommand{\ItemSpacing}{0mm}
\newcommand{\ParSpacing}{0mm}
\setenumerate[2]{itemsep={\ItemSpacing},parsep={\ParSpacing}}

\renewcommand{\vec}[1]{\mathbf{#1}}
\newcommand{\mat}[1]{\mathbf{#1}}

\newtheorem{Definition}{Definition}[section]

\newcommand\eps\varepsilon

\renewcommand\inf\infty

\renewcommand{\b}[1]{\textbf{#1}}

\newcommand{\x}{\vec{x}}

\newcommand{\X}{\mat{X}}

\newcommand{\R}{\mathbb{R}}

\newcommand{\vdelta}{\vec{\delta}}

\renewcommand{\rshift}{\texttt{>>}\text{ }}

\renewcommand{\sp}{\text{ }}

\DeclarePairedDelimiter\abs{\lvert}{\rvert}%

\DeclareMathOperator{\sign}{sign}
\DeclareMathOperator{\log2}{\text{log}_2}

\usepackage[all]{nowidow}

\setcopyright{acmlicensed}
\acmJournal{IMWUT}
\acmYear{2018} \acmVolume{2} \acmNumber{3} \acmArticle{93} \acmMonth{9} \acmPrice{15.00}\acmDOI{10.1145/3264903}

\received{May 2018}
\received{July 2018}
\received[accepted]{September 2018}

\newcommand{\mine}{\textsc{Sprintz}}
\newcommand{\minesp}{\textsc{Sprintz}\text{ }}
\newcommand{\fire}{\textsc{Fire}\text{ }}
\newcommand{\justfire}{\textsc{Fire}}

\begin{document}

\title{Sprintz: Time Series Compression for the Internet of Things}


\author{Davis Blalock}
\affiliation{
  \institution{Massachusetts Institute of Technology}
}
\email{dblalock@mit.edu}

\author{Samuel Madden}
\affiliation{
  \institution{Massachusetts Institute of Technology}
}
\email{madden@csail.mit.edu}

\author{John Guttag}
\affiliation{
  \institution{Massachusetts Institute of Technology}
}
\email{guttag@mit.edu}

\begin{abstract}

Thanks to the rapid proliferation of connected devices, sensor-generated time series constitute a large and growing portion of the world's data. Often, this data is collected from distributed, resource-constrained devices and centralized at one or more servers. A key challenge in this setup is reducing the size of the transmitted data without sacrificing its quality. Lower quality reduces the data's utility, but smaller size enables both reduced network and storage costs at the servers and reduced power consumption in sensing devices. A natural solution is to compress the data at the sensing devices. Unfortunately, existing compression algorithms either violate the memory and latency constraints common for these devices or, as we show experimentally, perform poorly on sensor-generated time series.

We introduce a time series compression algorithm that achieves state-of-the-art compression ratios while requiring less than 1KB of memory and adding virtually no latency. This method is suitable not only for low-power devices collecting data, but also for servers storing and querying data; in the latter context, it can decompress at over 3GB/s in a single thread, even faster than many algorithms with much lower compression ratios. A key component of our method is a high-speed forecasting algorithm that can be trained online and significantly outperforms alternatives such as delta coding.

Extensive experiments on datasets from many domains show that these results hold not only for sensor data but also across a wide array of other time series.

\end{abstract}



\maketitle

\renewcommand{\shortauthors}{D. Blalock et al.}

\section{Introduction} \label{sec:intro}






Thanks to the proliferation of smartphones, wearables, autonomous vehicles, and other connected devices, it is becoming common to collect large quantities of sensor-generated time series. Once this data is centralized in servers, many tools exist to analyze and create value from it \cite{spark, mapreduce, hive, hdfs, influxDB, druid, littleTable, openTSDB}. However, centralizing it can be challenging because of power constraints on the devices collecting the data. In particular, transmitting data wirelessly is extremely power-intensive---on a representative set of chips \cite{cc2540, cc2640}, transmitting data over Bluetooth Low Energy (BLE) costs tens of \textit{milliwatts}, while computing at full power costs only tens of \textit{microwatts}, and sitting idle costs close to 1 microwatt.

One strategy for reducing this power consumption is to extract information locally and only transmit summaries \cite{guttagAnanthaEEG, socialFMRI, respawnDB}. This can be effective in some cases, but requires both a predefined use case for which a summary is sufficient and an appropriate method of constructing this summary. Devising such a method can be a significant endeavor; e.g., Verma et al. \cite{guttagAnanthaEEG} conducted a research project to identify features in their data that would allow for accurate classification of brain activity.

A complementary and more general approach is to compress the data before transmitting it \cite{socialFMRI, lachCompress, sensorTransforms, iotSignals, iotCompressCrap}. This allows arbitrary subsequent analysis and does not require elaborate summary construction algorithms. Unfortunately, existing compression methods either 1) are only applicable for specific types of data, such as timestamps \cite{gorilla, berkeleyTreeDB, fastpfor}, audio \cite{flac, shorten, aac, vorbis} or EEG \cite{guttagAnanthaEEG, eegCS} recordings; or 2) use algorithms that are ill-suited to sensor-generated time series.

More specifically, existing methods (e.g., \cite{sax, tsCompressSmartGrid, ecgCompressLossy, apca, lz4, zstd, zlib, gzip, lemireSegmentation}) violate one or more of the following design requirements:

\begin{enumerate}
\item \b{Small block size}. On devices with only a few kilobytes of memory, it is not possible to buffer large amounts of data before compressing it. Moreover, even with more memory, buffering can add unacceptable latency; for example, a smartwatch transmitting nine axes of 8-bit motion data at 20Hz to a smartphone would need to wait $10000 / (9 \times 1 \times 20) = 56$ seconds to fill even a 10KB buffer. This precludes using this data for gesture recognition and would add tremendous user interface latency for step counting, activity recognition, or most other purposes.
\item \b{High decompression speed}. While the device collecting the data may not need to decompress it, it is desirable to have an algorithm that could also function well in a central database. This eliminates the need to transcode the data at the server and simplifies the application. In a database, time series workloads are not only read-heavy \cite{respawnDB, berkeleyTreeDB, influxDB}, but often necessitate materializing data (or downsampled versions thereof) for visualization, clustering, computing correlations, or other operations \cite{respawnDB}. At the same time, writing is often append-only \cite{gorilla, respawnDB}. As a result, decompression speed is paramount, while compression speed need only be fast enough to keep up with the rate of data ingestion.
\item \b{Lossless}. Given that time series are almost always noisy and often oversampled, it might not seem necessary to compress them losslessly. However, noise and oversampling 1) tend to vary across applications, and 2) can be addressed in an application-specific way as a preprocessing step. Consequently, instead of assuming that some level of downsampling or some particular smoothing will be appropriate for all data, it is better for the compression algorithm to preserve what it is given and leave preprocessing up to the application developer.
\end{enumerate}


The primary contribution of this work is \mine,
a compression algorithm for time series that offers state-of-the-art compression ratios and speed while also satisfying all of the above requirements. It requires $<$1KB of memory, can use blocks of data as small as eight samples, and can decompress at up to 3GB/s in a single thread. \mine's effectiveness stems from exploiting 1) temporal correlations in each variable's value and variance, and 2) the potential for parallelization across different variables, realized through the use of vector instructions. The main limitation of \minesp is that it operates directly only on integer time series. However, as we discuss in Section ~\ref{sec:floats}, straightforward preprocessing allows it to be applied to most floating point time series as well.

A key component of \mine's operation is a novel, vectorized forecasting algorithm for integers. This algorithm can simultaneously train online and generate predictions at close to the speed of \texttt{memcpy}, while significantly improving compression ratios compared to delta coding.

A second contribution is an empirical comparison of a range of algorithms currently used to compress time series, evaluated across a wide array of public datasets. We also make available code to easily reproduce these experiments, including the plots and statistical tests in the paper. To the best of our knowledge, this constitutes the largest public benchmark for time series compression.

The remainder of this paper is structured as follows. In Section~\ref{sec:problem}, we introduce relevant definitions, background, and details regarding the problem we consider. In Section~\ref{sec:relatedWork}, we survey related work and what distinguishes \mine. In Sections~\ref{sec:method} and \ref{sec:results}, we describe \minesp and evaluate it across a number of publicly available datasets. We also discuss when \minesp is advantageous relative to other approaches.



\section{Definitions and Background} \label{sec:problem}

Before elaborating upon how our method works, we introduce necessary definitions and provide relevant information regarding the problem being solved.

\subsection{Definitions}
\begin{Definition} \b{Sample.} A sample is a vector $\x \in \R^D$. $D$ is the sample's \b{dimensionality}. Each element of the sample is an integer represented using a number of bits $w$, the \b{bitwidth}. The bitwidth $w$ is shared by all elements.
\end{Definition}
\begin{Definition} \b{Time Series.} A time series $\X$ of length $T$ is a sequence of $\text{ }T$ samples, $\x_1,\ldots,\x_T$. All samples $\x_t$ share the same bitwidth $w$ and dimensionality $D$. If $D = 1$, $\X$ is called \b{univariate}; otherwise it is \b{multivariate}.
\end{Definition}
\begin{Definition} \b{Rows, Columns.} When represented in memory, we assume that each sample of a time series is one row and each dimension is one column. Because data arrives as samples and memory constraints may limit how many samples can be buffered, we assume that the data is stored in row-major order---i.e., such that each sample is stored contiguously.
\end{Definition}

\subsection{Hardware Constraints}

Many connected devices are powered by batteries or harvested energy \cite{bsnChallenges}. This results in strict power budgets and, in order to satisfy them, omission of certain functionality. In particular, many devices lack hardware support for floating point operations, SIMD (vector) instructions, and integer division. Moreover, they often have no more than a few kilobytes of memory, clocks in the tens of MHz at most, and 8-, 16-, or 32-bit processors instead of 64-bit \cite{cc2540, cc2640, quark}.

In contrast, we assume that the hardware used to decompress the data does not share these limitations. It is likely a modern x86 server with SIMD instructions, gigabytes of RAM, and a multi-GHz clock. However, because the amount of data it must store and query can be large, compression ratio and decompression speed are still important.

\subsection{Data Characteristics}

From a compression perspective, time series have four attributes uncommon in other data.

\begin{enumerate}
    \item \textbf{Lack of exact repeats}. In text or structured records, there are many sequences of bytes---often corresponding to words or phrases---that will exactly repeat many times. This makes dictionary-based methods a natural fit. In time series, however, the presence of noise makes exact repeats less common \cite{extract, epenthesis}.
    \item \textbf{Multiple variables}. Real-world time series often consist of multiple variables that are collected and accessed together. For example, the Inertial Measurement Unit (IMU) in modern smartphones collects three-dimensional acceleration, gyroscope, and magnetometer data, for a total of nine variables sampled at each time step. These variables are also likely to be read together, since each on its own is insufficient to characterize the phone's motion. 
    \item \textbf{Low bitwidth}. Any data collected by a sensor will be digitized into an integer by an Analog-to-Digital Converter (ADC). Nearly all ADCs have a precision of 32 bits or fewer \cite{digikeyADCs}, and typically 16 or fewer of these bits are useful. For example, even lossless audio codecs store only 16 bits per sample \cite{flac, shorten}. Even data that is not collected from a sensor can often be stored using six or fewer bits without loss of performance for many tasks \cite{epenthesis, mdlIntrinsic, sax}.
    \item \textbf{Temporal correlation}. Since the real world usually evolves slowly relative to the sampling rate, successive samples of a time series tend to have similar values. However, when multiple variables are present and samples are stored contiguously, this correlation is often present only with a lag---e.g., with nine IMU variables, every ninth value is similar. Such lag correlations violate the assumptions of most compressors, which treat adjacent bytes as the most likely to be related.
\end{enumerate}

Much of the reason \minesp outperforms existing methods is that it exploits or accounts for all of these characteristics, while existing methods do not.



\vspace{3mm}
\section{Related Work} \label{sec:relatedWork}

\minesp draws upon ideas from time series compression, time series forecasting, integer compression, general-purpose compression, and high-performance computing. From a technical perspective, \minesp is unusual or unique in its abilities to:
\begin{enumerate}
    \item Bit pack with extremely small block sizes
    \item Bit pack low-bitwidth integers effectively
    \item Efficiently exploit correlation between successive samples in \textit{multivariate} time series
    \item Naturally integrate both run-length encoding and bit packing
    \item Exploit vectorized hardware through forecaster, learning algorithm, and bit packing method co-design
\end{enumerate}
From an application persective, \minesp is distinct in that it enables higher-ratio lossless compression with far less memory and latency than competing methods.



\subsection{Compression of Time Series}

Most work on compressing time series has focused on lossy techniques. The most common approach is to approximate the data as a sequence of low-order polynomials \cite{swab, lemireSegmentation, tsCompressSmartGrid, iotCompressCrap, apca, paa}. An alternative, commonly seen in the data mining literature, is to discretize the time series using Symbolic Aggregate Approximation (SAX) \cite{sax} or its variations \cite{isax, isax2}. These approaches are designed to preserve enough information about the time series to support indexing or specific data mining algorithms (e.g. \cite{isax, fastShapelet, hotSax}), rather than to compress the time series \textit{per se}. As a result, they are extremely lossy; a hundred-sample time series might be compressed into one or two bytes, depending on the exact discretization parameters. 


For audio time series specifically, there are a large number of lossy codecs \cite{vorbis, shorten, aac, opus}, as well as a small number of lossless \cite{flac, alac} codecs. In principle, some of these could be applied to non-audio time series. However, modern codecs make such strong assumptions about the possible numbers of channels, sampling rates, bit depths, or other characteristics that it is infeasible to use them on non-audio time series.

Many fewer algorithms exist for lossless time series compression. For floating-point time series, the only algorithm of which we are aware is that of the Gorilla database \cite{gorilla}. This method XORs each value with the previous value to obtain a diff, and then bit packs the diffs. In contrast to our approach, it assumes that time series are univariate and have 64-bit floating-point elements. 

For lossless compression of integer time series (including timestamps), existing approaches include directly applying general-purpose compressors \cite{respawnDB, openTSDB, chronicleDB, kairosDB, druid}, (double) delta encoding and then applying an integer compressor \cite{influxDB, gorilla}, or predictive coding and byte-packing \cite{akumuli}. These approaches can work well, but tend to offer both less compression and less speed than \mine.

\subsection{Compression of Integers}


The fastest methods of compressing integers are generally based on bit packing---i.e., using at most $b$ bits to represent values in $\{0, 2^b-1\}$, and storing these bits contiguously \cite{bbp, pfor, fastpfor}. Since $b$ is determined by the largest value that must be encoded, naively applying this method yields limited compression. To improve it, one can encode fixed-size blocks of data at a time, so that $b$ can be set based on the largest values in a block instead of the whole dataset \cite{kGamma, pfor, fastpfor}. A further improvement is to ignore the largest few values when setting $b$ and store their omitted bits separately \cite{pfor, fastpfor}.

\minesp bit packing differs significantly from existing methods in two ways. First, it compresses much smaller blocks of samples. This reduces its throughput as compared to, e.g., \cite{fastpfor}, but significantly improves compression ratios (c.f. Section~\ref{sec:results}). This is because large values only increase $b$ for a few samples instead of for many. Second, \minesp is designed for 8 and 16-bit integers, rather than 32 or 64-bit integers. Existing methods are often inapplicable to lower-bitwidth data (unless converted to higher-bitwidth data) thanks to strong assumptions about bitwidth and data layout.

A common \cite{flac, shorten} alternative to bit packing is Golomb coding \cite{golomb}, or its special case Rice coding \cite{rice}. The idea is to assume that the values follow a geometric distribution, often with a rate constant fit to the data. 

Both bit packing and Golomb coding are bit-based methods in that they do not guarantee that encoded values will be aligned on byte boundaries. When this is undesirable, one can employ byte-based methods such as 4-Wise Null Suppression \cite{kGamma}, LEB128 \cite{dwarf}, or Varint-G8IU \cite{varintG8IU}. These methods reduce the number of bytes used to store each sample by encoding in a few bits how many bytes are necessary to represent its value, and then encoding only that many bytes. Some, such as Simple8B \cite{simple8b} and SIMD-GroupSimple \cite{groupSimd}, allow fractional bytes to be stored while preserving byte alignment for groups of samples. 


\subsection{General-Purpose Compression}
A reasonable alternative to using a time series compressor would be to apply a general-purpose compression algorithm, possibly after delta coding or other preprocessing. Thanks largely to the development of Asymmetric Numeral Systems (ANS) \cite{ans} for entropy coding, general purpose compressors have advanced greatly in recent years. In particular, Zstd \cite{zstd}, Brotli \cite{brotli}, LZ4 \cite{lz4} and others have attained speed-compression tradeoffs significantly better than traditional methods such as GZIP \cite{gzip}, LZO \cite{lzo}, etc. However, these methods have much higher memory requirements than \minesp and, empirically, often do not compress as well and/or decompress as quickly.


\subsection{Predictive Filtering}



For numeric data such as time series, there are four types of predictive coding commonly in use: predictive filtering \cite{png}, delta coding \cite{fastpfor, bbp}, double-delta coding \cite{influxDB, gorilla}, and XOR-based encoding \cite{gorilla}. In predictive filtering, each prediction is a linear combination of a fixed number of recent samples. This can be understood as an autoregressive model or the application of a Finite Impulse Response (FIR) filter. When the filter is learned from the data, this is termed ``adaptive filtering.'' Many audio compressors use some form of adaptive filtering \cite{shorten, flac, aac}.

Delta coding is a special case of predictive filtering where the prediction is always the previous value. Double-delta coding, also called delta-delta coding or delta-of-deltas coding, consists of applying delta coding twice in succession. XOR-based encoding is similar to delta coding, but replaces subtraction of the previous value with the XOR operation. This modification is often desirable for floating-point data \cite{gorilla}.

Our forecasting method can be understood as a special case of adaptive filtering. While adaptive filtering is a well-studied mathematical problem in the signal processing literature, we are unaware of a practical algorithm that attains speed within an order of magnitude of our own. I.e., our method's primary novelty is as a vectorized \textit{algorithm} for fitting and predicting multivariate time series, rather than as a mathematical \textit{model} of multivariate time series. That said, it does incorporate different modeling assumptions than other compression algorithms for time series in that it reduces the model to one parameter and omits a bias term. 





\section{Method} \label{sec:method}


To describe how \minesp works, we first provide an overview of the algorithm, then discuss each of its component in detail.


\subsection{Overview}

\minesp is a bit packing-based predictive coder. It consists of four components:
\begin{enumerate}
\item \b{Forecasting.} \minesp employs a forecaster to predict each sample based on previous samples. It encodes the difference between the next sample and the predicted sample, which is typically closer to zero than the next sample itself.
\item \b{Bit packing.} \minesp then bit packs the errors as a ``payload'' and prepends a header with sufficient information to invert the bit packing.
\item \b{Run-length encoding.} If a block of errors is all zeros, \minesp waits for a block in which some error is nonzero and then writes out the number of all-zero blocks instead of the (otherwise empty) payload.
\item \b{Entropy coding.} \minesp Huffman codes the headers and payloads.
\end{enumerate}

These components are run on blocks of eight samples (motivated in Section~\ref{sec:bitpacking}), and can be modified to yield different compression-speed tradeoffs. Concretely, one can 1) skip entropy coding for greater speed and 2) choose between delta coding and our online learning method as forecasting algorithms. The latter is slightly slower but often improves compression.

We chose these steps since they allow for high speed and exploit the characteristics of time series. Forecasting leverages the high correlation of successive samples to reduce the entropy of the data. Run-length encoding allows for extreme compression in the (common) scenario that there is no change in the data---e.g., a user's smartphone may be stationary for many hours while the user is asleep. Our method of bit packing exploits temporal correlation in the variability of the data by using the same bitwidth for points that are within the same block. Huffman coding is not specific to time series but has low memory requirements and improves compression ratios.





\newcommand{\err}{\texttt{err}}
\newcommand{\nbits}{\texttt{nbits}}
\newcommand{\packed}{\texttt{packed}}
\newcommand{\buff}{$\texttt{buff}$}
\newcommand{\bytes}{\texttt{bytes}}
\newcommand{\payload}{\texttt{payload}}
\newcommand{\f}{\texttt{f}}
\newcommand{\fore}{\texttt{forecaster}}
\newcommand{\self}{\texttt{self}}


An overview of how \minesp compresses one block of samples is shown in Algorithm~\ref{algo:compress}. In lines \ref{line:bodyStart}-\ref{line:encPredictEnd}, \minesp predicts each sample based on the previous sample and any state stored by the forecasting algorithm. For the first sample in a block, the previous sample is the last element of the previous block, or zeros for the initial block. In lines \ref{line:eachColStart}-\ref{line:bodyEnd}, \minesp determines the number of bits required to store the largest error in each column and then bit packs the values in that column using that many bits. (Recall that each column is one variable of the time series). If all columns require 0 bits, \minesp continues reading in blocks until some error requires $>$0 bits (lines \ref{line:rleLoopStart}-\ref{line:rleLoopEnd}). At this point, it writes out a header of all 0s and then the number of all-zero blocks. Finally, it writes out the number of bits required by each column in the latest block as a header, and the bit packed data as a payload. Both header and payload are compressed with Huffman coding.

\begin{algorithm}[h]
\caption{encodeBlock($\{\x_1, \ldots, \x_B \}, \fore$)}
\label{algo:compress}
\begin{algorithmic}[1]

\State{Let \buff\sp be a temporary buffer}

\For {$i \leftarrow 1,\ldots,B$} \COMMENTT {For each sample} \label{line:bodyStart}
    \State{$ \hat{\x}_i \leftarrow $ $\fore$.predict$(\x_{i-1})$}
    \State{$ \err_i \leftarrow \x_i - \hat{\x}_i  $}
    \State{$\fore$.train($\x_{i-1}$, $\x_i$, $\err_i$)} \label{line:encPredictEnd}
\EndFor
\For {$j \leftarrow 1,\ldots,D$} \COMMENTT {For each column} \label{line:eachColStart}
    \State{$ \nbits_j \leftarrow \max_i\{ $requiredNumBits$(\err_{ij}) \} $}
    \State{$ \packed_j \leftarrow $ bitPack$(\{\err_{1j},\ldots,\err_{Bj} \},\text{ }\nbits_j) $}  \label{line:bodyEnd}
\EndFor

\LineComment{Run-length encode if all errors are zero}
\If{$\nbits_j$ \texttt{==} $0$, $1 \le j \le D$}
    \Repeat  \COMMENT{Scan until end of run} \label{line:rleLoopStart}
        \State{Read in another block and run lines \ref{line:bodyStart}-\ref{line:bodyEnd} }
    \Until {$\exists_j [\nbits_j \neq 0 ]$} \label{line:rleLoopEnd}
    \State{Write $D$ $0$s as headers into \buff}
    \State{Write number of all-zero blocks as payload into \buff}
    \State{Output huffmanCode(\buff)}
\EndIf

\State{Write $\nbits_j$, $j = 1,\ldots,D$ as headers into \buff}
\State{Write $\packed_j$, $j = 1,\ldots,D$ as payload into \buff}
\State{Output huffmanCode(\buff)}

\end{algorithmic}
\end{algorithm}


\minesp begins decompression (Algorithm~\ref{algo:decomp}) by decoding the Huffman-coded bitstream into a header and a payload. Once decoded, these two components are easy to separate since the header is always first and of fixed size. If the header is all 0s, the payload indicates the length of a run of zero errors. In this case, \minesp runs the predictor until the corresponding number of samples have been predicted. Since the errors are zero, the forecaster's predictions are the true sample values. In the nonzero case, \minesp unpacks the payload using the number of bits specified for each column by the header.

\begin{algorithm}[h]
\caption{decodeBlock(\bytes, $B$, $D$, $\fore$)}
\label{algo:decomp}
\begin{algorithmic}[1]

\State{$\nbits$, $\payload$ $\leftarrow$ huffmanDecode(\bytes, $B$, $D$) }

\If{$\nbits_j$ \texttt{==} $0$ $\forall j$} \COMMENT{Run-length encoded}
    \State{$\texttt{numblocks} \leftarrow $ readRunLength()}
    \For {$i \leftarrow 1,\ldots,(B $ $\cdot$ \texttt{numblocks})}
        \State{$ \x_i \leftarrow $ $\fore$.predict$(\x_{i-1})$}
        \State{Output $\x_i$}
    \EndFor
\Else \COMMENT{Not run-length encoded}
\For {$i \leftarrow 1,\ldots,B$}
    \State{$ \hat{\x}_i \leftarrow $ $\fore$.predict$(\x_{i-1})$}
    \State{$ \err_i \leftarrow $unpackErrorVector$(i$, \nbits, \payload$) $}
    \State{$ \x_i \leftarrow \err_i + \hat{\x}_i  $}
    \State{Output $\x_i$}
    \State{$\fore$.train($\x_{i-1}$, $\x_i$, $\err_i$)}
\EndFor
\EndIf
\end{algorithmic}
\end{algorithm}

\subsection{Forecasting}

\minesp forecasting can use either delta coding or \justfire \text{ } (Fast Integer REgression), a novel online forecasting algorithm we introduce.

\subsubsection{Delta Coding}

Forecasting with delta coding consists of predicting each sample $\x_i$ to be equal to the previous sample $\x_{i-1}$, where $\x_{0} \triangleq \vec{0}$. This method is stateless given $\x_{i-1}$ and is extremely fast. It is particularly fast when combined with run-length encoding, since it yields a run of zero errors if and only if the data is constant. This means that decompression of runs requires only copying a fixed vector, with no additional forecasting or training. Moreover, when answering queries, one can sometimes avoid decompression entirely---e.g., one can compute the max of all samples in the run by computing the max of only the first value.

\subsubsection{FIRE}

Forecasting with \fire is slightly more expensive than delta coding but often yields better compression.
The basic idea of \fire is to model each value as a linear combination of a fixed number of previous values and learn the coefficients of this combination. Specifically, we learn an autoregressive model of the form:
\begin{align}
    x_i = a x_{i-1} + b x_{i-2} + \eps_i
\end{align}
where $x_i$ denotes the value of some variable at time step $i$ and $\eps_i$ is a noise term.

Different values of $a$ and $b$ are suitable for different data characterisics. If $a = 2$, $b = -1$, we obtain double-delta coding, which extrapolates linearly from the previous two points and works well when the time series is smooth. If $a = 1$, $b = 0$, we recover delta coding, which models the data as a random walk. If $a = \frac{1}{2}$, $b = \frac{1}{2}$, we predict each value to be the average of the previous two values, which is optimal if the $x_i$ are i.i.d. Gaussians. In other words, these cases are appropriate for successively noisier data.


The reason \fire is effective is that it learns online what the best coefficients are for each variable.
To make prediction and learning as efficient as possible, \justfire \text{} restricts the coefficients to lie within a useful subspace. Specifically, we exploit the observation that all of the above cases can be written as:
\begin{align}
    x_i = x_{i-1} + \alpha x_{i-1} - \alpha x_{i-2} + \eps_i
\end{align}
for $\alpha \in [-\frac{1}{2}, 1]$. Letting $\delta_i \triangleq x_i - x_{i-1}$ and subtracting $x_{i-1}$ from both sides, this is equivalent to
\begin{align}
    \delta_i &= \alpha \delta_{i-1} + \eps_i
\end{align}
This means that we can capture all of the above cases by predicting the next delta as a rescaled version of the previous delta. This requires only a single addition and multiplication, and reduces the learning problem to that of finding a suitable value for a single parameter.


To train and predict using this model, we use the functions shown in Algorithm~\ref{algo:xff}. First, to initialize a \fire forecaster, one must specify three values: the number of columns $D$, the learning rate $\eta$, and the bitwidth $w$ of the integers stored in the columns. Internally, the forecaster also maintains an accumulator for each column (line~\ref{line:counter}) and the difference (delta) between the two most recently seen samples (line~\ref{line:deltas}). The accumulator is a scaled version of the current $\alpha$ value with a bitwidth of $2w$. It enables fast updates of $\alpha$ with greater numerical precision than would be possible if modifying $\alpha$ directly. The accumulators and deltas are both initialized to zeros. 


\begin{algorithm}[h]
\caption{FIRE\_Forecaster Class} \label{algo:xff}
\begin{algorithmic}[1]

\Function{Init}{$D$, $\eta$, $w$} \label{line:xffCtor}
\State $\self$.learnShift $\leftarrow \lg(\eta)$
\State $\self$.bitWidth $\leftarrow w$ \COMMENT{8-bit or 16-bit}
\State $\self$.accumulators $\leftarrow $ zeros($D$) \label{line:counter}
\State $\self$.deltas $\leftarrow $ zeros($D$) \label{line:deltas}
\EndFunction

\Function{Predict}{$\x_{i-1}$} \label{line:xffPredict}
\State $\texttt{alphas} \leftarrow \self$.accumulators \rshift $\self$.learnShift
\State $\hat{\vdelta} \leftarrow$ (\texttt{alphas} $\odot$ $\self$.deltas) $\rshift \self$.bitWidth
\RETURN $\x_{i-1} + \hat{\vdelta}$
\EndFunction

\Function{Train}{$\x_{i-1}$, $\x_{i}$, $\err_i$} \label{line:xffTrain}
\State $\texttt{gradients} \leftarrow {-\sign(\err_i)} \odot \self$.deltas
\State $\self$.accumulators $\leftarrow$ $\self$.accumulators $-$ $\texttt{gradients}$
\State $\self$.deltas $\leftarrow \x_{i} - \x_{i-1}$
\EndFunction

\end{algorithmic}
\end{algorithm}

To predict, the forecaster first derives the coefficient $\alpha$ for each column based on the accumulator. By right shifting the accumulator $\log2(\eta)$ bits, the forecaster obtains a learning rate of $2^{-\log2(\eta)} = \eta$. It then estimates the next deltas as the elementwise product (denoted $\odot$) of these coefficients and the previous deltas. It predicts the next sample to be the previous sample plus these estimated deltas.

Because all values involved are integers, the multiplication is done using twice the bitwidth $w$ of the data type---e.g., using 16 bits for 8 bit data. The product is then right shifted by an amount equal to the bit width. This has the effect of performing a fixed-point multiplication with step size equal to $2^{-w}$.

The forecaster trains by performing a gradient update on the L1 loss between the true and predicted samples. I.e., given the loss:
\begin{align}
    \mathcal{L}(x_i, \hat{x}_{i}) = \abs{x_i - \hat{x}_i}
    = \abs{x_i - (x_{i-1} + \frac{\alpha}{2^w} \cdot \delta_{i-1})} \\
    = \abs{\delta_{i} - \frac{\alpha}{2^w} \cdot \delta_{i-1}}
\end{align}
for one column's value $x_i = \x_{ij}$ for some $j$ and coefficient $\alpha$, the gradient is:
\begin{align}
        \frac{\partial }{\partial \alpha} \abs{\delta_{i} - \frac{\alpha}{2^w} \cdot \delta_{i-1}}
&= \begin{cases}
        -{2^{-w}}\vdelta_{i-1} & \x_{i} > \hat{\x}_{i} \\
        {2^{-w}}\vdelta_{i-1} & \x_{i} \le \hat{\x}_{i}
\end{cases} \\
&= -\sign(\eps) \cdot {2^{-w}}\vdelta_{i-1} \\
&\propto -\sign(\eps) \cdot \vdelta_{i-1}
\end{align}
where we define $\eps \triangleq \x_{i} - \hat{\x}_{i}$ and ignore the $2^{-w}$ as a constant that can be absorbed into the learning rate. In all experiments reported here, we set the learning rate to $\frac{1}{2}$. This value is unlikely to be ideal for any particular dataset, but preliminary experiments showed that it consistently worked reasonably well. 

In practice, \fire differs from the above pseudocode in three ways. First, instead of computing the coefficient for each sample, we compute it once at the start of each block. Second, instead of performing a gradient update after each sample, we average the gradients of all samples in each block and then perform one update. Finally, we only compute a gradient for every other sample, since this has little or no effect on the accuracy and slightly improves speed.

\begin{figure*}[t]
\begin{center}
    \includegraphics[width=\textwidth]{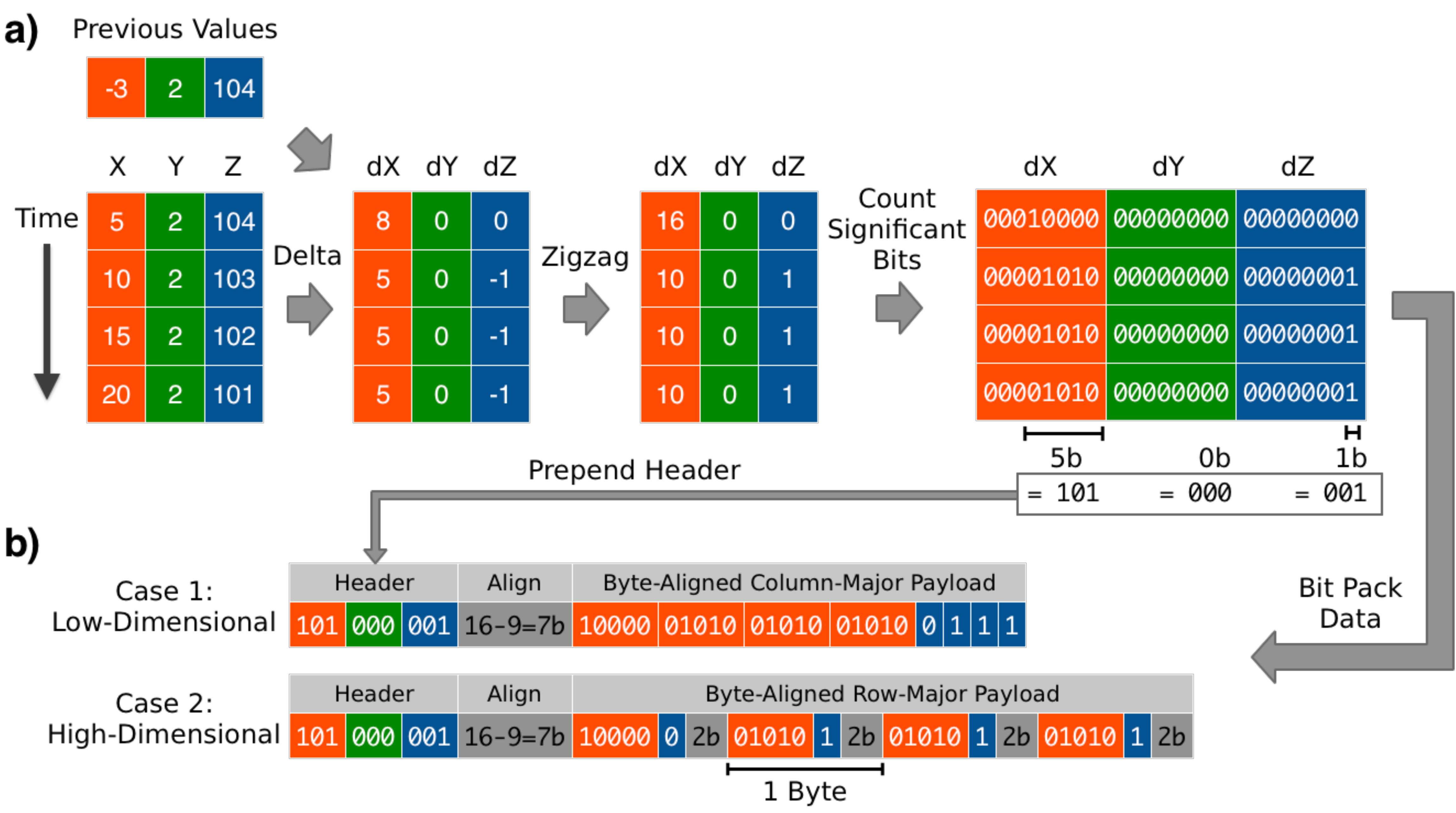}
    \caption{Overview of \minesp using a delta coding predictor.\textit{ a)} Delta coding of each column, followed by zigzag encoding of resulting errors. The maximum number of significant (nonzero) bits is computed for each column. \textit{b)} These numbers of bits are stored in a header, and the original data is stored as a (byte-aligned) payload, with leading zeros removed. When there are few columns, each column's data is stored contiguously. When there are many columns, each row is stored contiguously, possibly with padding to ensure alignment on a byte boundary.}
    \label{fig:overview}
    \vspace{-5mm}
\end{center}
\end{figure*}

\subsection{Bit Packing} \label{sec:bitpacking}

An illustration of \mine's bit packing is given in Figure~\ref{fig:overview}. The prediction errors from delta coding or \fire are zigzag encoded \cite{zigzag} and then the minimum number of bits required is computed for each column. Zigzag encoding is an invertible transform that interleaves positive and negative integers such that each integer is represented by twice its absolute value, or twice its absolute value minus one for negative integers. This makes all values nonnegative and maps integers farther from zero to larger numbers.

Given the zigzag encoded errors, the number of bits $w^\prime$ required in each column can be computed as the bitwidth minus the fewest leading zeros in any of that column's errors. E.g., in Figure~\ref{fig:overview}a, the first column's largest encoded value is 16, represented as \texttt{00010000}, which has three leading zeros. This means that we require $w^\prime = 8 - 3 = 5$ bits to store the values in this column. One can find this value by ORing all the values in a column together and then using a built-in function such as GCC's $\texttt{\_\_builtin\_clz}$ to compute the number of leading zeros in a single assembly instruction (c.f. \cite{fastpfor}). This optimization motivates our use of zigzag encoding to make all values nonnegative.

Once the number of bits $w^\prime$ required for each column is known, the zigzag-encoded errors can be bit packed. First, \minesp writes out a header consisting of $D$ unsigned integers, one for each column, storing the bitwidths. Each integer is stored in $\log2(w)$ bits, where $w$ is the bitwidth of the data. Since there are $w+1$ possible values of $w^\prime$ (including 0), width $w-1$ is treated as a width of $w$ by both the encoder and decoder. E.g., 8-bit data that could only be compressed to 7 bits is both stored and decoded with a bitwidth of 8.

After writing the headers, \minesp takes the appropriate number of low bits from each element and packs them into the payload. When there are few columns, all the bits for a given column are stored contiguously (i.e., column-major order). When there are many columns, the bits for each \textit{sample} are stored contiguously (i.e., row-major order). In the latter case, up to seven bits of padding are added at the end of each row so that all rows begin on a byte boundary. This means that the data for each column begins at a fixed bit offset within each row, facilitating vectorization of the decompressor. The threshold for choosing between the two formats is a sample width of $32$ bits.

The reason for this threshold is as follows. Because the block begins in row-major order and we seek to reconstruct it the same way, the row-major bit packing case is more natural. For small numbers of columns, however, the row padding can significantly reduce the compression ratio. Indeed, for univariate 8-bit data, it makes compression ratios greater than 1 impossible. This gives rise to the column-major case; using a block size of eight samples and column-major order, each column's data always falls on a byte boundary without any padding. The downside of this approach is that both encoder and decoder must transpose the block. However, for up to four 8-bit columns or two 16-bit columns, this can be done quickly using SIMD shuffling instructions.\footnote{For recent processors with AVX-512 instructions, one could double these column counts, but we refrain from assuming that these instructions will be available.} This gives rise to the cutoff of 32 bit sample width for choosing between the formats.

As a minor bit packing optimization, one can store the headers for two or more blocks contiguously, so that there is one group of headers followed by one group of payloads. This allows many headers to share one set of padding bits between the headers and payload. Grouping headers does not require buffering more than one block of raw input, but it does require buffering the appropriate number of blocks of compressed output. In addition to slightly improving the compression ratio, it also enables more headers to be unpacked with a given number of vector instructions in the decompressor. Microbenchmarks show up to $10\%$ improvement in decompression speed as the number of blocks in a group grows towards eight. However, we use groups of only two in all reported experiments to ensure that our results tend towards pessimism and are applicable under even the most extreme buffer size constraints.

\subsection{Entropy Coding}

We entropy code the bit packed representation of each block using Huff0, an off-the-shelf Huffman coder \cite{fse}. This encoder treats individual bytes as symbols, regardless of the bitwidth of the original data. We use Huffman coding instead of Finite-State Entropy \cite{fse} or an arithmetic coding scheme since they are slower, and we never observed a meaningful increase in compression ratio.

The benefit of adding Huffman coding to bit packing stems from bit packing's inability to optimally encode individual bytes. For a given packed bitwidth $w$, bit packing models its input as being uniformly distributed over an interval of size $2^{w}$. Appropriately setting $w$ allows it to exploit the similar variances of nearby values, but does not optimally encode individual values (unless they truly are uniformly distributed within the interval). Huffman coding is complementary in that  it fails to capture relationships between nearby bytes but optimally encodes individual bytes.

We Huffman code after bit packing, instead of before, for two reasons. First, doing so is faster. This is because the bit packed block is usually shorter than the original data, so less data is fed to the Huffman coding routines. These routines are slower than the rest of \mine, so minimizing their input size is beneficial. Second, this approach increases compression. Bit packed bytes benefit from Huffman coding, but Huffman coded bytes do not benefit from bit packing, since they seldom contain large numbers of leading zeros. This absence of leading zeros is unsurprising since Huffman codes are not byte-aligned and use ones and zeros in nearly equal measure.



\subsection{Vectorization}

Much of \mine's speed comes from vectorization. For headers, the fixed bitwidths for each field and fixed number of fields allows for packing and unpacking with a mix of vectorized byte shuffles, shifts, and masks. For payloads, delta (de)coding, zigzag (de)coding, and \fire all operate on each column independently, and so naturally vectorize. Because the packed data for all rows is the same length and aligned to a byte boundary (in the high-dimensional case), the decoder can compute the bit offset of each column's data one time and then use this information repeatedly to unpack each row. In the low-dimensional case, all packed data fits in a single vector register which can be shuffled/masked appropriately for each possible number of columns. This is possible since there are at most four columns in this case. On an \texttt{x86} machine, bit packing and unpacking can be accelerated with the \texttt{pext} and \texttt{pdep} instructions, respectively.

\section{Experimental Results} \label{sec:results}


To asses \mine's effectiveness, we compared it to a number of state-of-the art compression algorithms on a large set of publicly available datasets. All of our code and raw results are publicly available on the \minesp website.\footnote{https://smarturl.it/sprintz} This website also contains additional experiments, as well as documentation of both our code and experimental setups. All experiments use a single thread on a 2013 Macbook Pro with a 2.6GHz Intel Core i7-4960HQ processor.

All reported timings and throughputs are the best of ten runs. We use the best, rather than average, since this is 1) desirable in the presence of the non-random, purely additive noise characteristic of microbenchmarks, and, 2) consequently, a best practice in microbenchmarking \cite{lemireMicrobenchmarks}. The best values are nearly always within 10\% of the averages.

\subsection{Datasets}

\begin{itemize}[leftmargin=4mm]
\itemsep0mm
\item \textbf{UCR} \cite{ucrTimeSeries} --- The UCR Time Series Archive is a repository of 85 univariate time series datasets from various domains, commonly used for benchmarking time series algorithms. Because each dataset consists of many (often short) time series, we concatenate all the time series from each dataset to form a single longer time series. This is to allow dictionary-based methods to share information across time series (instead of compressing each in isolation). To mitigate artificial jumps in value from the end of one time series to the start of another, we linearly interpolate five samples between each pair.
\item \textbf{PAMAP} \cite{pamap} --- The PAMAP dataset consists of inertial motion and heart rate data from wearable sensors on subjects performing everyday actions. It has 31 variables, most of which are accelerometer and gyroscope readings.
\item \textbf{MSRC-12} \cite{msrc} --- The MSRC-12 dataset consists of 80 variables of (x, y, z, depth) positions of human joints captured by a Microsoft Kinect. The subjects performed various gestures one might perform when interacting with a video game.
\item \textbf{UCI Gas} \cite{uci_gas} --- This dataset consists of 18 columns of gas concentration readings and ground truth concentrations during a chemical experiment.
\item \textbf{AMPDs} \cite{ampds} --- The Almanac of Minutely Power Datasets describes electricity, water, and natural gas consumption recorded once per minute for two years at a single home. 
\end{itemize}


For datasets stored as delimited files, we first parsed the data into a contiguous, numeric array and then dumped the bytes as a binary file. Before obtaining any timing results, we first load each dataset into main memory.
Because the datasets are provided as floating point values (despite most reflecting analog-to-digital converter output that was originally integer-valued), we quantized them into integers before operating on them. We did so by linearly rescaling them such that the largest and smallest values corresponded to the largest and smallest values representable with the number of bits tested---e.g., 0 and 255 for 8 bits---and then applying the floor function. Note that this is the worst case scenario for our method since it maximizes the number of bits required to represent the data.

For multivariate datasets, we allowed all methods but our own to operate on the data one variable at a time; i.e., instead of interleaving values for every variable, we store all values for each variable contiguously. This corresponds to allowing them an unlimited buffer size in which to store incoming data before compressing it. We allow these ideal conditions in order to ensure that our results for existing methods err towards optimism and to eliminate buffer size as a lurking variable.


\subsection{Comparison Algorithms}

\begin{itemize}[leftmargin=4mm]
\itemsep0mm
\item \textbf{SIMD-BP128} \cite{fastpfor} --- The fastest known method of compressing integers.
\item \textbf{FastPFOR} \cite{fastpfor} --- An algorithm similar to SIMD-BP128, but with better compression ratios.
\item \textbf{Simple8b} \cite{simple8b} --- An integer compression algorithm used by the popular time series database InfluxDB \cite{influxDB}.
\item \textbf{Snappy} \cite{snappy} --- A general-purpose compression algorithm developed by Google and used by InfluxDB, KairosDB \cite{kairosDB}, OpenTSDB \cite{openTSDB}, RocksDB \cite{rocksDB}, the Hadoop Distributed File System \cite{hdfs} and numerous other projects.
\item \textbf{Zstd} \cite{zstd} --- Facebook's state-of-the-art general purpose compression algorithm. It is based on LZ77 and entropy codes using a mix of Huffman coding and Finite State Entropy (FSE) \cite{fse}. It is available in RocksDB \cite{rocksDB}.
\item \textbf{LZ4} \cite{lz4} --- A widely-used general-purpose compression algorithm optimized for speed and based on LZ77. It is used by RocksDB and ChronicleDB \cite{chronicleDB}.
\item \textbf{Zlib} \cite{zlib} --- A popular implementation of the DEFLATE \cite{deflate} dictionary coder, which also underlies gzip \cite{gzip}.

\end{itemize}

For Zlib and Zstd, we use a compression level of 9 unless stated otherwise. This level heavily prioritizes compression ratio at the expense of increased compression time. We use it to improve the results for these methods in experiments in which compression time is not penalized.

We also assess three variations of \mine, corresponding to different speed/ratio tradeoffs:
\begin{enumerate}
    \item \textbf{\texttt{SprintzFIRE+Huf}}. The full algorithm described in Section~\ref{sec:method}.
    \item \textbf{\texttt{SprintzFIRE}}. Like \texttt{SprintzFIRE+Huf}, but without Huffman coding.
    \item \textbf{\texttt{SprintzDelta}}. Like \texttt{SprintzFIRE}, but with delta coding instead of \fire as the forecaster.
\end{enumerate}



\subsection{Compression Ratio}

In order to rigorously assess the compression performance of both \minesp and existing algorithms, it is desirable to evaluate each on a large corpus of time series from heterogeneous domains. Consequently, we use the UCR Time Series Archive \cite{ucrTimeSeries}. This corpus contains dozens of datasets and is almost universally used for evaluating time series classification and clustering algorithms in the data mining community.

\begin{figure}[h]
\begin{center}
    \includegraphics[width=10cm]{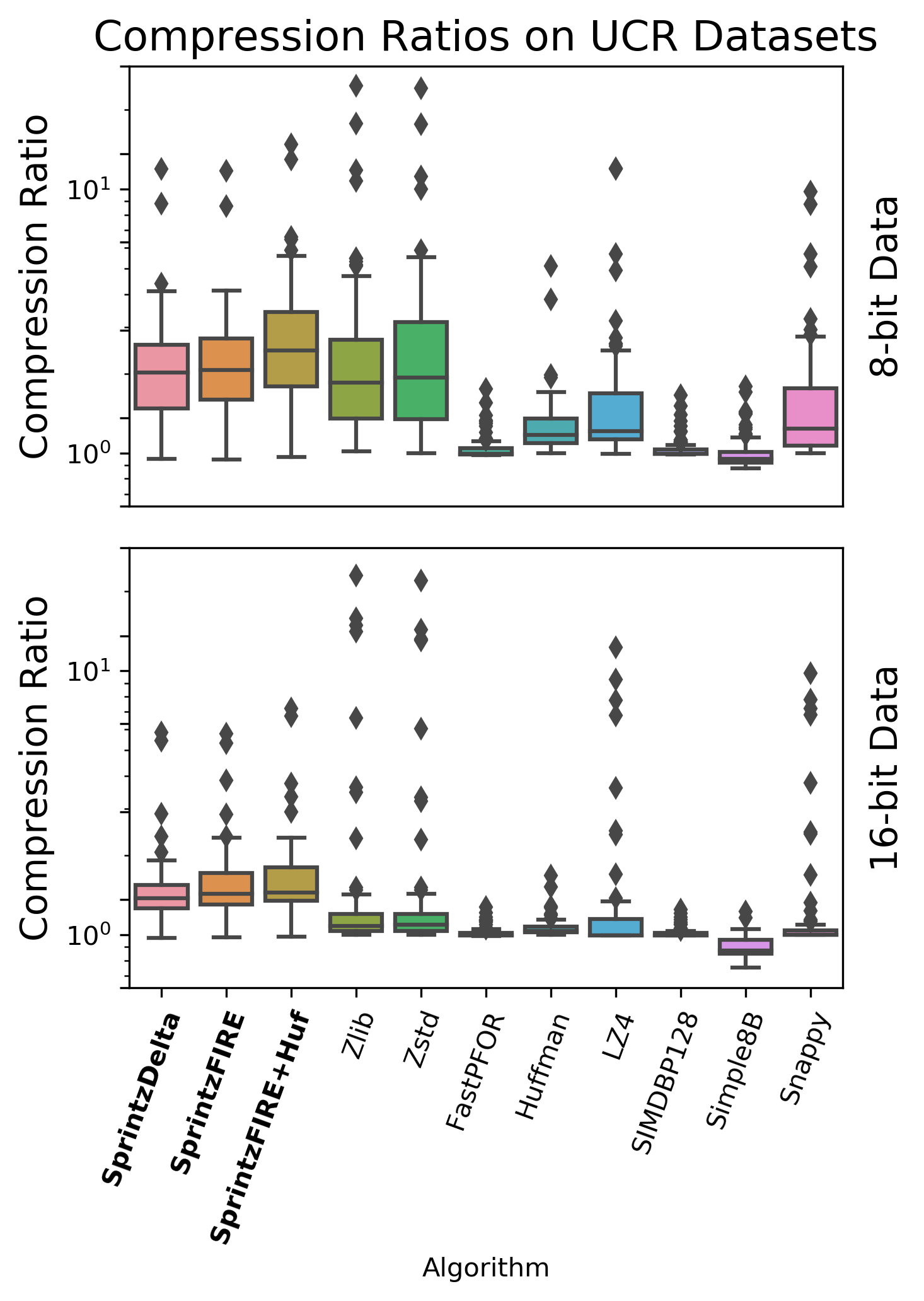}
    \caption{Boxplots of compression performance of different algorithms on the UCR Time Series Archive. Each boxplot captures the distribution of one algorithm across all 85 datasets.}
    \label{fig:ratioBox}
\end{center}
\end{figure}

The distributions of compression ratios on these datasets for the above algorithms are shown in in Figure~\ref{fig:ratioBox}. \minesp exhibits consistently strong performance across almost all datasets. High-speed codecs such as Snappy, LZ4, and the integer codecs (FastPFOR, SIMDBP128, Simple8B) hardly compress most datasets at all.

Perhaps counter-intuitively, 8-bit data tends to yield higher compression ratios than 16-bit data. This is a product of the fact that the number of bits that are ``predictable'' is roughly constant. I.e., suppose that an algorithm can correctly predict the four most significant bits of a given value; this enables a 2:1 compression ratio in the 8-bit case, but only a 16:12 = 4:3 ratio in the 16-bit case. Interestingly, the fact that trailing bits tend to be too noisy to compress also suggests that one could use a lower bitwidth with little loss of information. 

To assess \mine's performance statistically, we use a Nemenyi test \cite{nemenyiTest} as recommended in \cite{cdDiagrams}. This test compares the mean rank of each algorithm across all datasets, where the highest-ratio algorithm is given rank 1, the second-highest rank 2, and so on. The intuition for why this test is desirable is that it not only accounts for multiple hypothesis testing in making pairwise comparisons, but also prevents a small number of large or highly compressible datasets from dominating the results.

\begin{figure}[h]
\begin{center}
    \includegraphics[width=10cm]{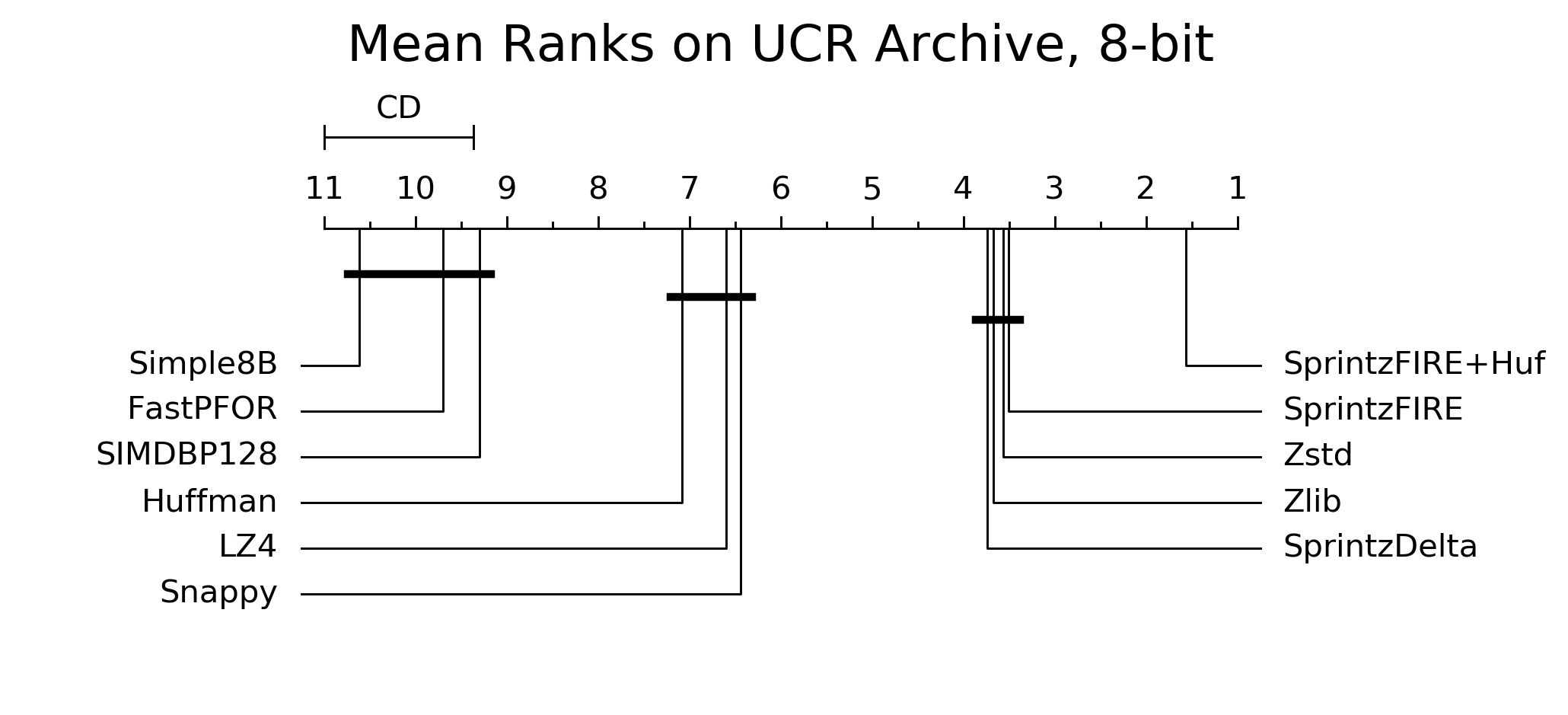}
    \includegraphics[width=10cm]{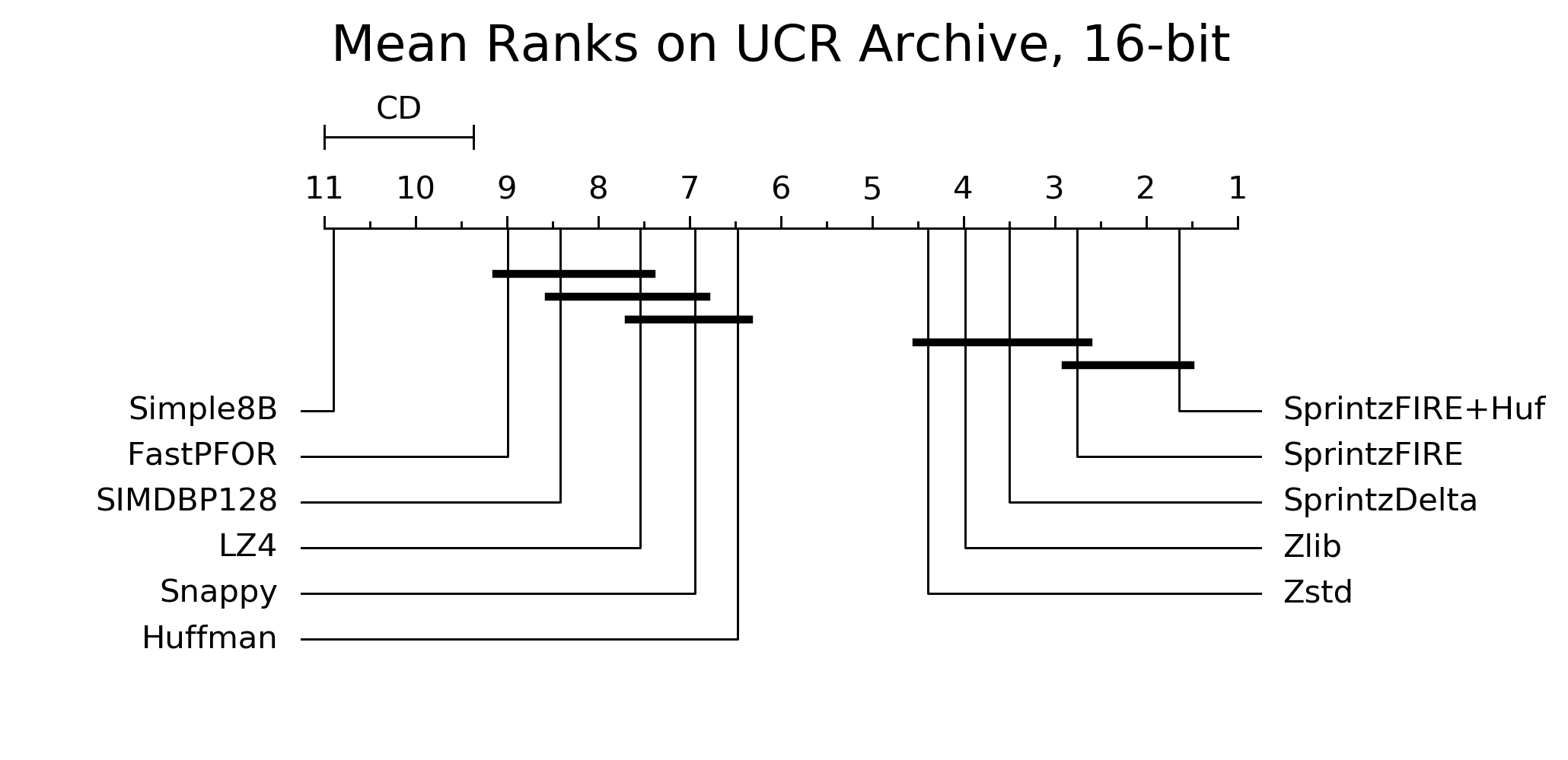}
    \caption{Compression performance of different algorithms on the UCR Time Series Archive. The x-axis is the mean rank of each method, where rank 1 on a given dataset had the highest ratio. Methods joined with a horizontal black line are not statistically significantly different.}
    \label{fig:ratioCD}
    \vspace{-5mm}
\end{center}
\end{figure}

The results of the Nemenyi test are shown in the Critical Difference Diagrams \cite{cdDiagrams} in Figure~\ref{fig:ratioCD}. These diagrams show the mean rank of each algorithm on the x-axis and join methods that are not statistically significantly different with a horizontal line. \minesp on high compression settings is significantly better than any existing algorithm. On lower settings, it is still as effective as the best current methods (Zlib and Zstd).





In addition to this overall comparison, it is important to assess whether \fire improves performance compared to delta coding. Since this is a single hypothesis with matched pairs, we assess it using a Wilcoxon signed rank test. This yields p-values of .0094 in the 8-bit case and 4.09e-12 in the 16-bit case. As a more interpretable measure, \fire obtains better compression on 51 of 85 datasets using 8 bits and 74 of 85 using 16. These results suggest that \fire is generally beneficial on 8-bit data but even more beneficial on 16-bit data.

To understand why 16-bit data benefits more, we examined datasets where \fire gives differing benefits in the two cases. The difference most commonly occurs when the data is highly compressible with just delta coding. With 8 bits and $\sim$$4\times$ compression, the forecaster's task is effectively to guess whether the next delta is -1, 0, or 1 given a current delta drawn from this same set. The Bayes error rate is high for this problem, and \justfire's attempt to learn adds variance compared to the delta coding approach of always predicting 0. In contrast, with 16 bits, the deltas span many more values and retain continuity that \fire can exploit.





\subsection{Decompression Speed} \label{sec:decomp_speed}

To systematically assess the speed of \minesp, we ran it on time series with varying numbers of columns and varying levels of compressibility. Because real datasets have a fixed and limited number of columns, we ran this experiment on synthetic data. Specifically, we generated a synthetic dataset of 100 million random values uniformly distributed across the full range of those possible for the given bitwidth. This data is incompressible and thus provides a worst-case estimate of \mine's speed (though in practice, we find that the speed is largely consistent across levels of compressibility).

We compressed the data with \minesp set to treat it as if it had 1 through 80 columns. Numbers that do not evenly divide the data size result in \minesp \texttt{memcpy}-ing the trailing bytes.

While using this synthetic data cannot tell us anything about \mine's compression ratio, it is suitable for throughput measurement. This is because both \mine's sequence of instructions executed and memory access patterns are effectively independent of the data distribution---\mine's core loop has no conditional branches and \minesp's memory accesses are always sequential. Moreover, it exhibits throughputs on real data matching or slightly exceeding the numbers below for the corresponding number of columns (c.f. Figure~\ref{fig:tradeoff_success}). 

\begin{figure}[h]
\begin{center}
    \includegraphics[width=10cm]{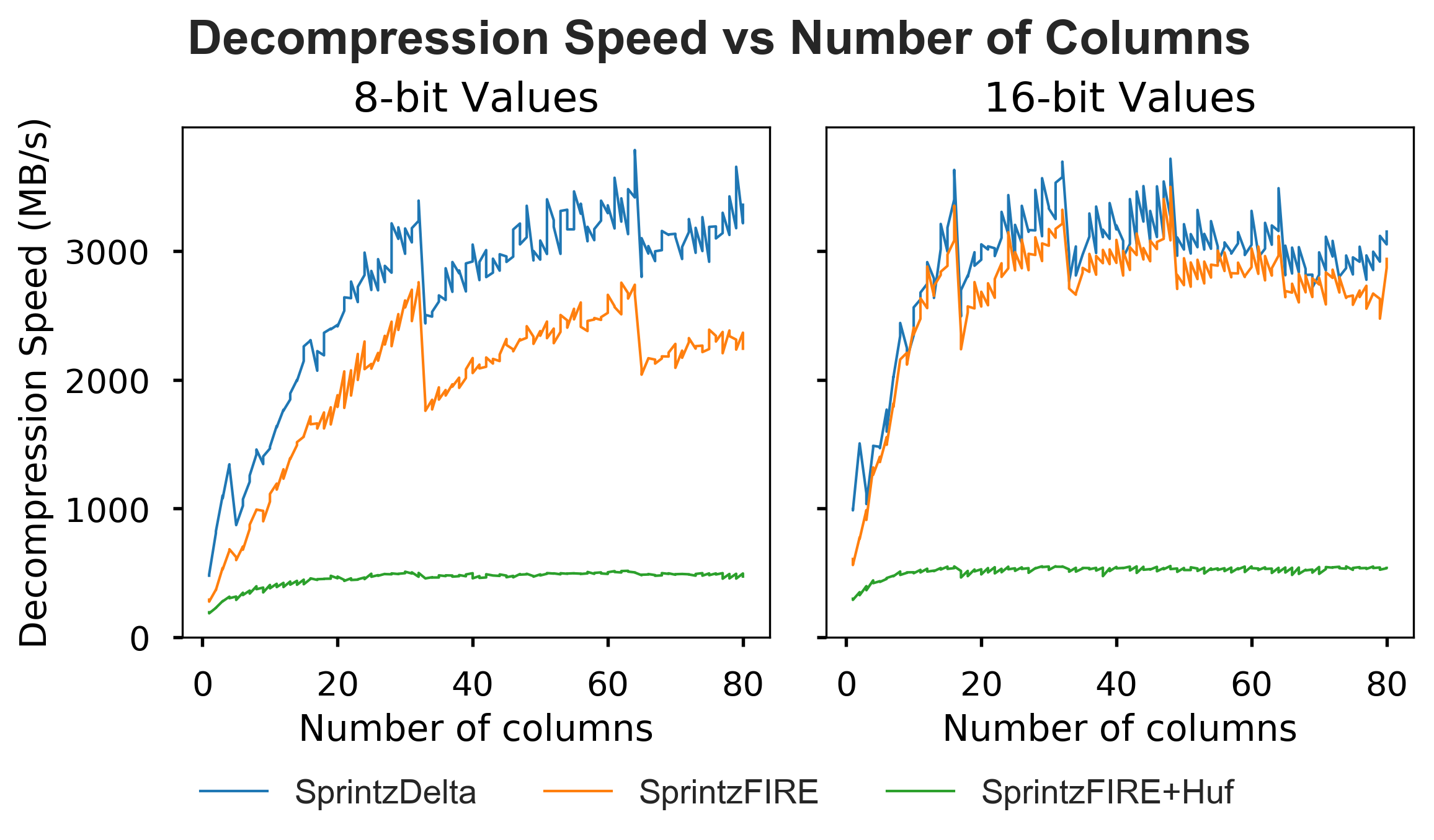}
    \caption{\minesp becomes faster as the number of columns increases and as the width of each sample approaches multiples of 32B (on a machine with 32B vector registers). }
    \label{fig:ndims_vs_decomp_speed}
\end{center}
\end{figure}

As shown in Figure~\ref{fig:ndims_vs_decomp_speed}, \minesp becomes faster as the number of columns increases and as the number of columns approaches multiples of 32 for 8-bit data or 16 for 16-bit data. These values correspond to the 256-bit width of a SIMD register on the machine used for testing. There is small but consistent overhead associated with using \fire over delta coding, but both approaches are extremely fast. Without Huffman coding, \minesp decompresses at multiple GB/s once rows exceed $\sim$16B. With Huffman coding, the other components of \minesp are no longer the bottleneck and \minesp consistently decompresses at over 500MB/s. Note that we omit comparison to other algorithms in this section since their speed varies with compressibility, not number of columns; see Section~\ref{sec:whenSprintz} for a direct comparison. Further note that the speed's dependence on number of columns is not an artifact of more columns yielding larger blocks of data. The limiting factor is serial dependence between decoding one sample and predicting the next one; this is accelerated by having wider samples that fill a vector register, but not by having longer blocks.

\subsection{Compression Speed} \label{sec:comp_speed}

\begin{figure}[h]
\begin{center}
    \includegraphics[width=10cm]{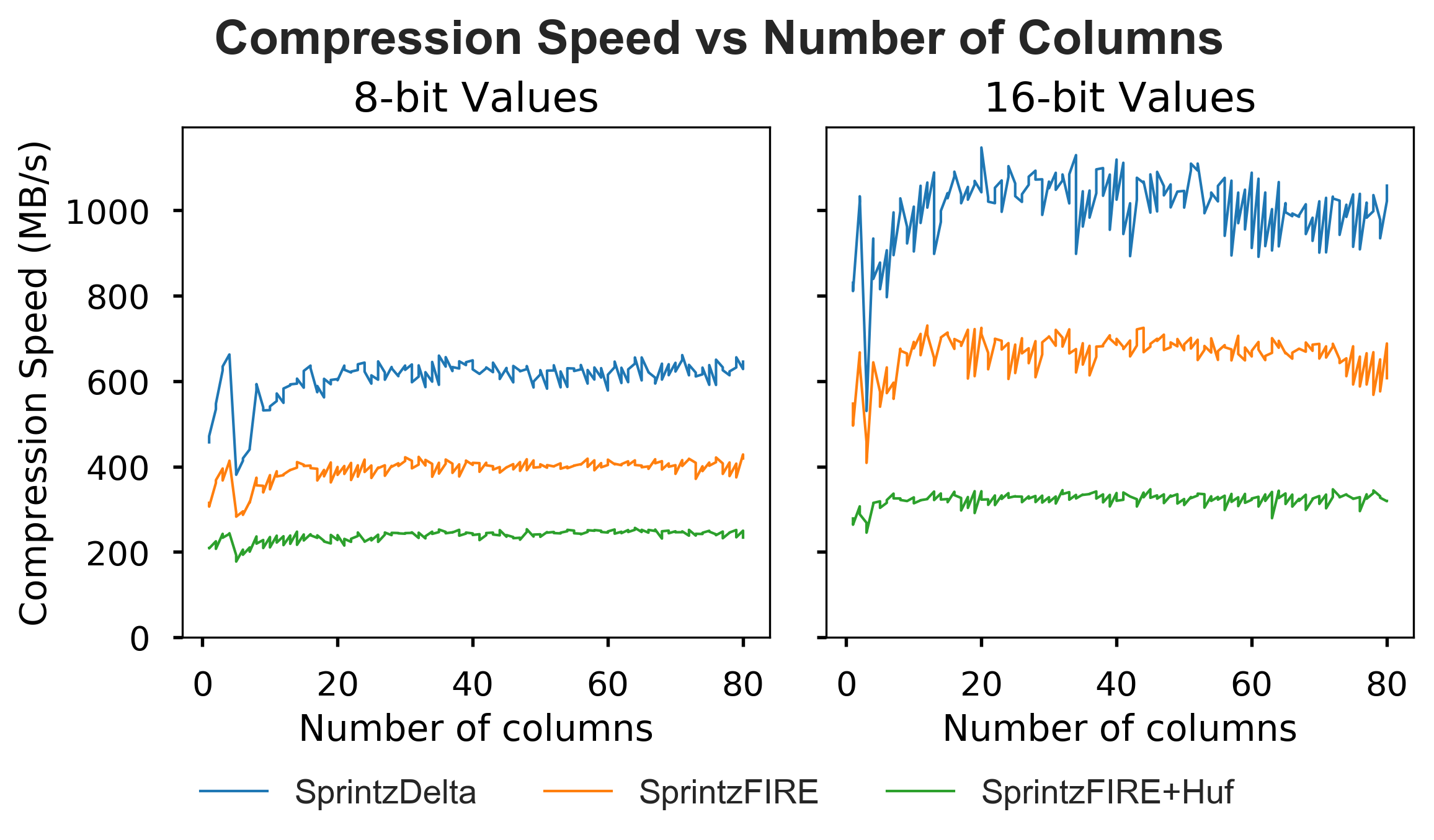}
    \caption{\minesp compresses at hundreds of MB/s even in the slowest case: its highest-ratio setting with incompressible 8-bit data. On lower settings with 16-bit data, it can exceed 1GB/s.}
    \label{fig:ndims_vs_comp_speed}
\end{center}
\end{figure}

It is important that \mine's compression speed be fast enough to keep up with the rate of data ingestion. We measured \mine's compression speed using the same methodology as decompression speed. As shown in Figure~\ref{fig:ndims_vs_comp_speed}, \mine \text{} compresses 8-bit data at over 200MB/s on the highest-ratio setting and 600MB/s on the fastest setting. These numbers are roughly 50\% larger on 16-bit data. We refrained from vectorizing this prototype implementation because 1) 200MB/s is already fast enough to run in real time even if every thread were fed data from its own gigabit network connection, and 2) low-power devices often lack vector instructions, so the measured speeds are more indicative of the rate at which these devices could compress (if scaled to the appropriate clock frequency). We again omit comparison to other compressors for the same reason as in the previous section.

The dips after 4 columns in 8-bit data and 2 columns in 16-bit data correspond to the switch from column-major bit packing to rowmajor bit packing.

\subsection{FIRE Speed}

\begin{figure}[h]
\begin{center}
    \includegraphics[width=10cm]{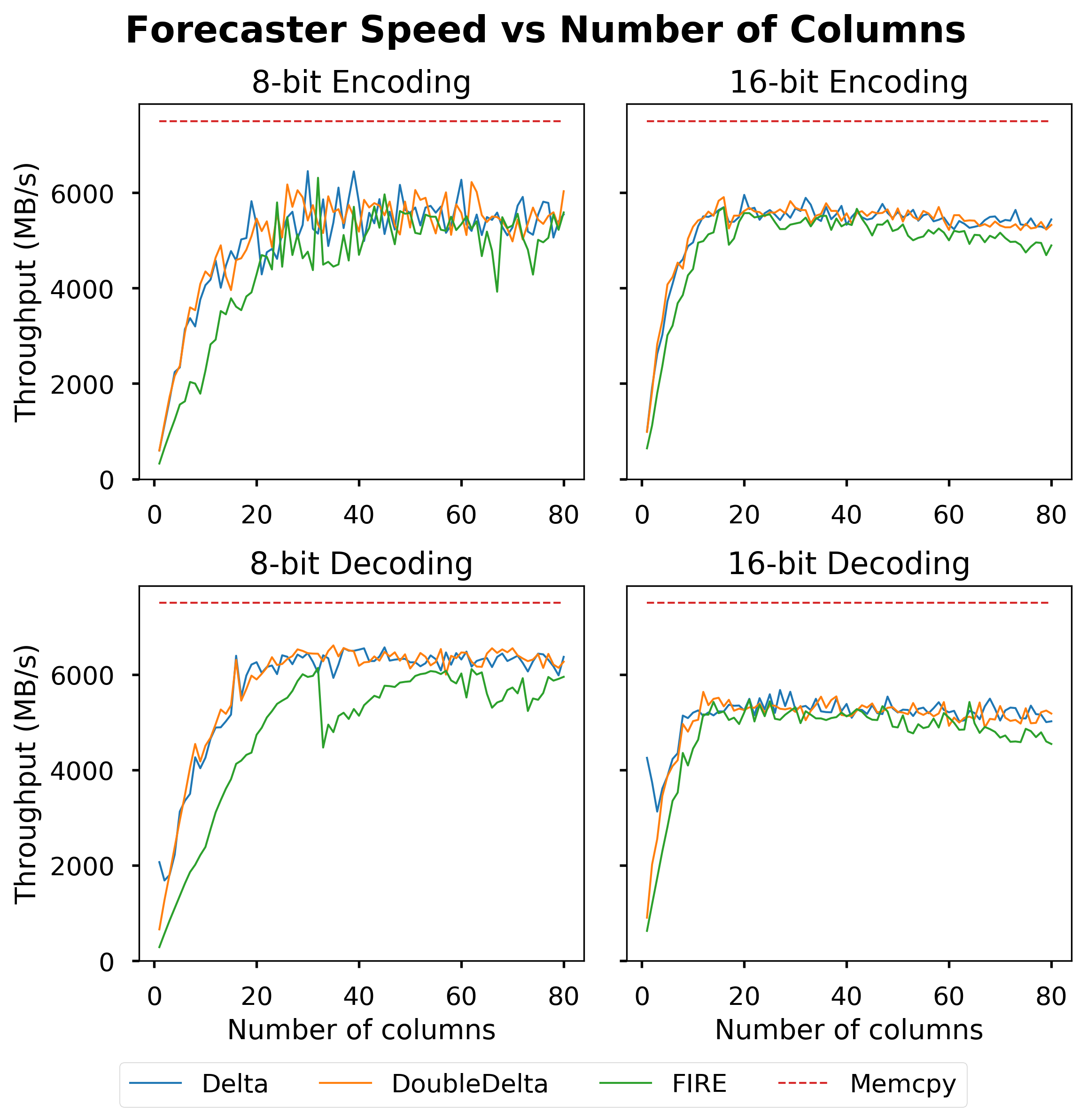}
    \caption{\fire is nearly as fast as delta and double delta coding. For a moderate number of columns, it runs at 5-6GB/s on a machine with 7.5GB/s \texttt{memcpy} speed.}
    \label{fig:ndims_vs_preproc_speed}
\end{center}
\end{figure}

To characterize the speed of the \fire we repeated the above throughput experiments with both it and two other predictors commonly seen in the literature: delta and double delta coding. As shown in Figure~\ref{fig:ndims_vs_preproc_speed}, \fire can encode at up to 5GB/s and decode at up to 6GB/s. This is nearly the same speed as the competing methods and close to the 7.5 GB/s speed of \texttt{memcpy} on the tested machine. Note that ``encode'' and ``decode'' here mean converting raw samples to errors and reconstructing samples from sequences of errors, respectively. These operations do not change the data size, but are the subroutines run in the \minesp compressor and decompressor. The reason that there is less discrepancy between delta and \fire encoding in isolation versus when embedded in \minesp compression (Figure~\ref{fig:ndims_vs_comp_speed}) is that, in this experiment, the implementations are vectorized.

\subsection{When to Use Sprintz} \label{sec:whenSprintz}


The above experiments provide a characterization of \mine's speed and a statistically meaningful assessment of its compression ratio in general. However, because one often wants to obtain the best results on a particular type of data, it is helpful to know when \minesp is likely to work well or poorly.

Regarding speed, \minesp is most desirable when there are many variables to be compressed. We have found that the speed is largely insensitive to compression ratio, so the results in Sections~\ref{sec:decomp_speed}~and~\ref{sec:comp_speed} offer a good estimate of the speed one could expect on similar hardware. The exception to this is if the data contains long runs of constants (or constant slopes if using \justfire). In this case, the decompression speed approaches the speed of \texttt{memcpy} for \texttt{SprintzDelta} or the speed of \fire for \texttt{SprintzFIRE} and \texttt{SprintzFIRE+Huf}.



Regarding compression ratio, the dominant requirement is that the data must have relatively strong correlations between consecutive values. This occurs when the sampling rate is fast relative to the time scale over which the measured quantity changes---the typical case when one seeks reasonably high-quality measurements. When these correlations are absent, predictive filtering (with only a two-component filter) has little value. Indeed, it can even be counterproductive. Consider the case of data that has an isolated nonzero value every few samples---e.g., the sequence $\{0, -1, 0, 0\}$. When delta coded, this yields $\{0, -1, 1, 0\}$, which requires an extra bit for \minesp bit packing. In general, \minesp has to pay the cost of abrupt changes twice---once when they happen, and once when they ``revert'' to the previous level.

\begin{figure}[t]
\begin{center}
    \includegraphics[width=10cm]{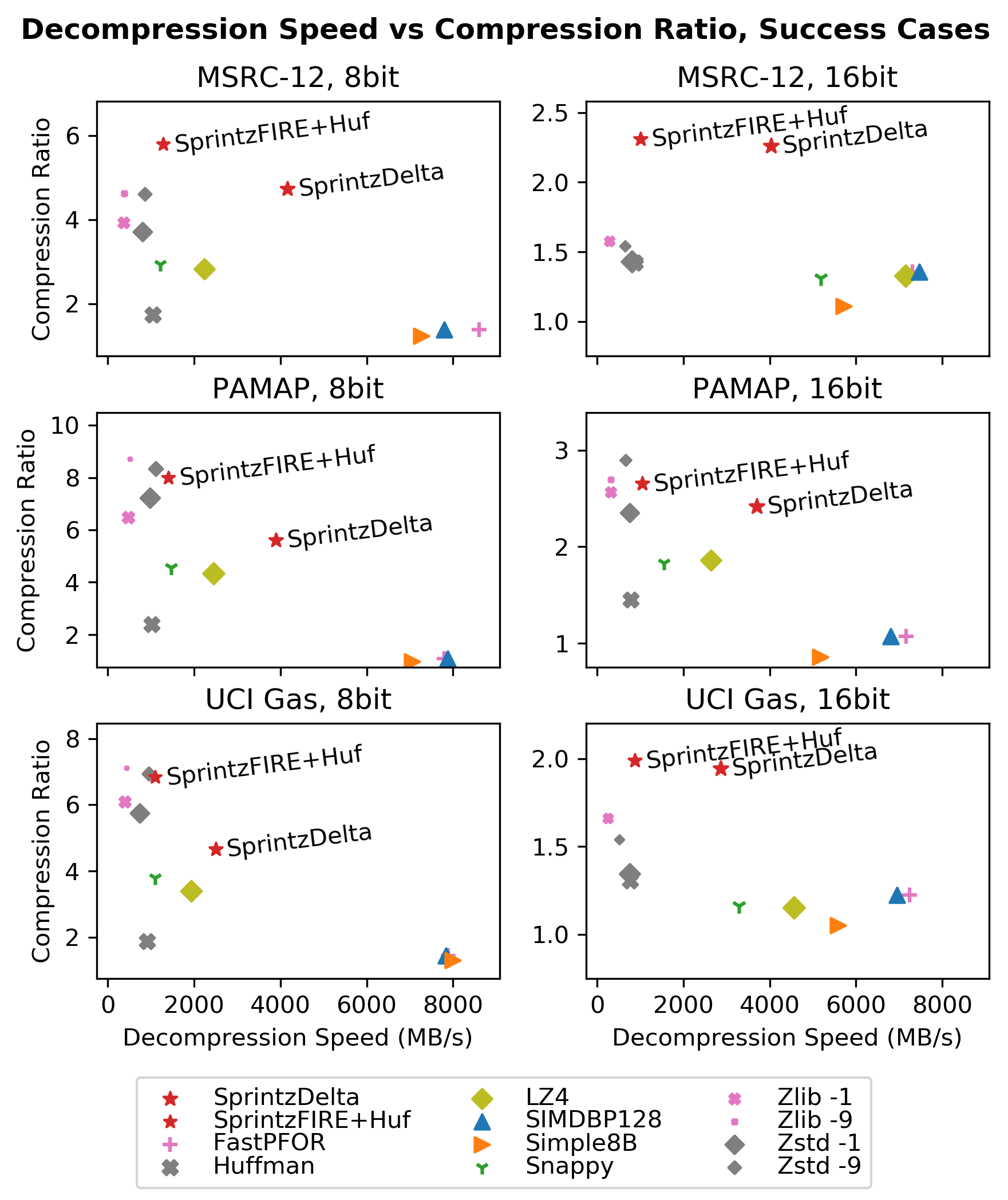}
    \caption{\minesp achieves excellent compression ratios and speed on relatively slow-changing time series with many variables.}
    \label{fig:tradeoff_success}
\end{center}
    \vspace{-5mm}
\end{figure}

Another specific case in which \minesp is undesirable is when the data distribution tends to switch between discrete states. For example, in electricity consumption data, an appliance tends to use little or no electricity when it is off and a relatively constant amount when it is on. Switches between these states are expensive for \mine, and predictive filtering offers little benefit on sequences of samples that are already almost constant. \minesp can still achieve reasonably good compression in this situation, but dictionary-based compressors will likely perform better. This is because they suffer no penalty from state changes, and runs of constants are their best-case input in terms of both ratio and speed. Their ratio benefits because they can often run-length encode the number of repeated values, and their speed benefits because they can decode runs at memory speed by \texttt{memcpy}-ing the repeated values.

As an illustration of when \minesp is and is not preferable, we ran it and the comparison algorithms on several real-world datasets with differing characteristics. In Figure~\ref{fig:tradeoff_success}, we use the MSRC-12, PAMAP and UCI Gas datasets. These datasets contain time series that change slowly relative to the sampling rate and have 80, 31, and 18 variables, respectively. \minesp achieves an excellent ratio-speed tradeoff on all three datasets, and the highest compression of any method \textit{even on its lowest-compression setting} on the MSRC-12 dataset.

\begin{figure}[h]
\begin{center}
    \includegraphics[width=10cm]{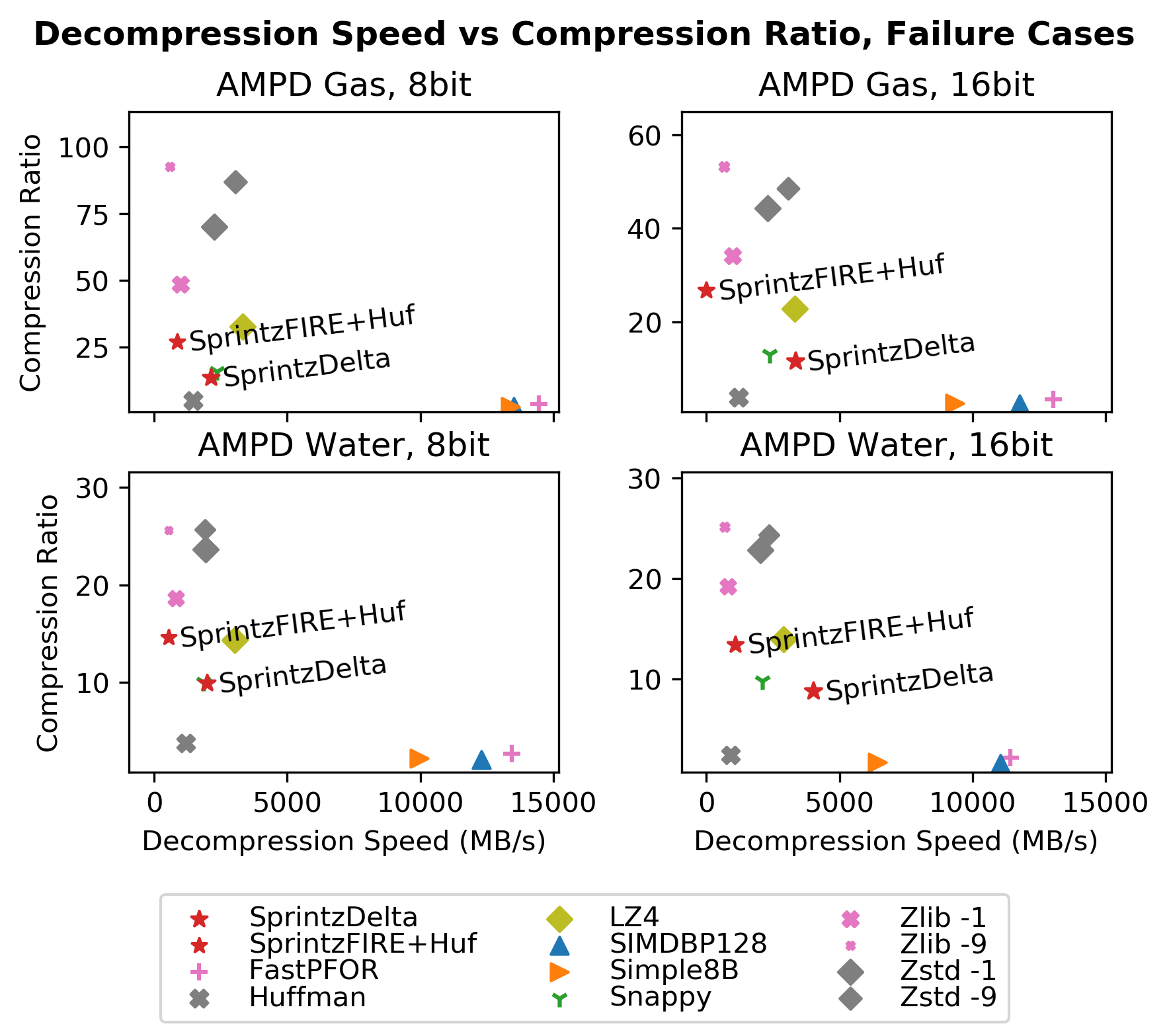}
    \caption{\minesp is less effective than other methods when the time series has large, abrupt changes and few variables.}
    \label{fig:tradeoff_failure}
\end{center}
\end{figure}

In contrast, \minesp performs poorly on the AMPD Gas and AMPD Water datasets (Figure~\ref{fig:tradeoff_failure}). These datasets chronicle the natural gas and water consumption of a house over a year, and often switch between discrete states and/or have isolated nonzero values. They also have only three and two variables, respectively. \minesp achieves more than $10\times$ compression, but dictionary-based methods such as Zstd and LZ4 achieve even greater compression, while also decompressing faster.

\subsection{Generalizing to Floats} \label{sec:floats}

While floating-point values are not the focus of this work, it is possible to apply \minesp to floats by first quantizing the floating-point data. The downside of doing this is that, because floating-point numbers are not uniformly distributed along the real line, such quantization is lossy. To assess the degree of loss, we carried out an experiment to measure the error induced when quantizing real data. Note that this experiment does not assess whether \minesp is the \textit{best} means of compressing floats---it merely suggests that using integer compressors like \minesp as lossy floating-point compressors is reasonable and could be a fruitful avenue for future work. 

We assessed the magnitude of typical quantization errors by quantizing the UCR time series datasets. Specifically, we linearly offset and rescaled the time series in each dataset such that the minimum and maximum values in any time series correspond to $(0, 255)$ for 8-bit quantization or $(0, 65535)$ for 16-bit quantization. We then obtained the quantized data by applying the floor function to this linearly transformed data.

\begin{figure}[h]
\begin{center}
    \includegraphics[width=10cm]{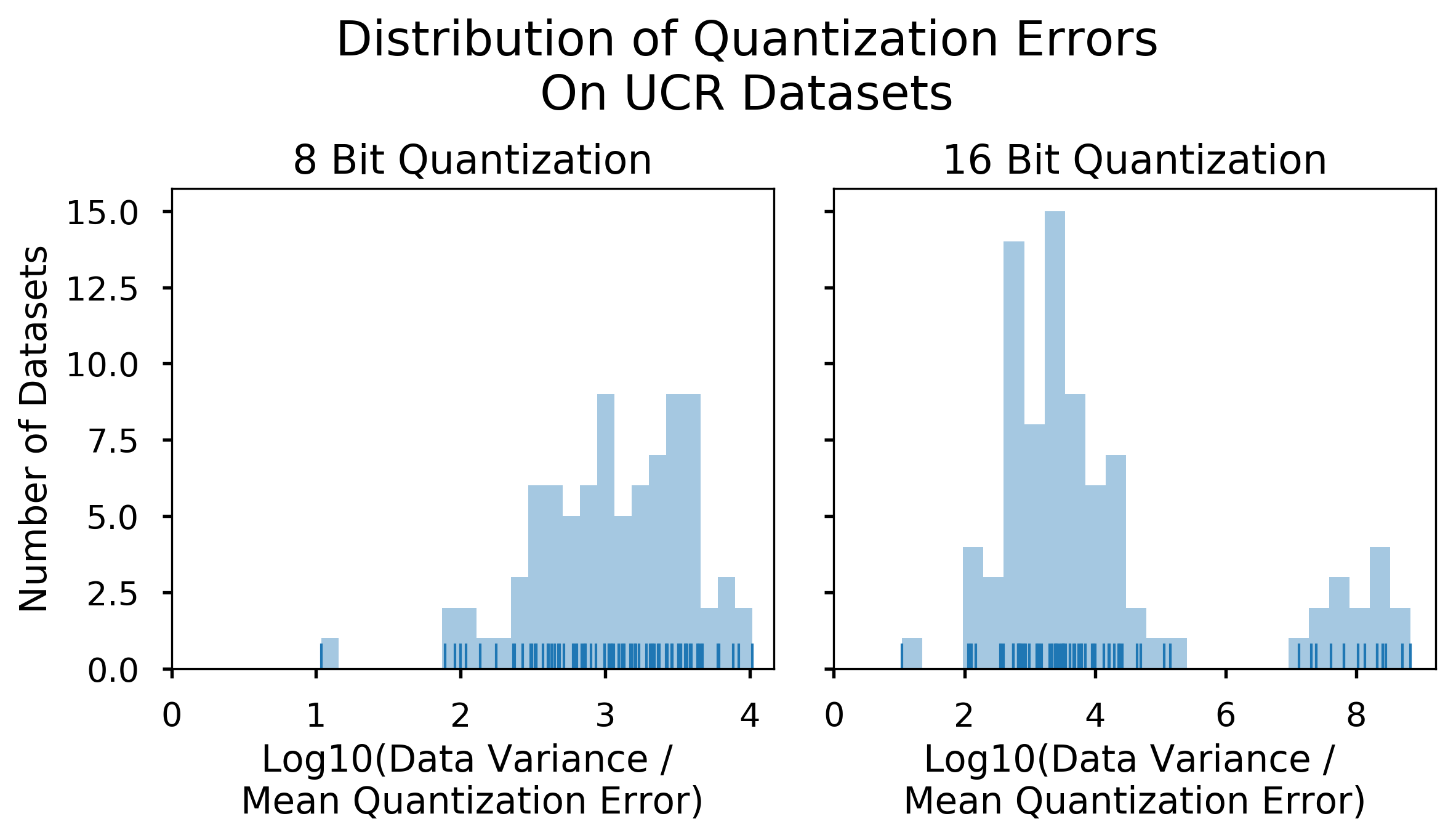}
    \caption{Quantizing floating-point time series to integers introduces error that is orders of magnitude smaller than the variance of the data. Even with 8 bits, quantization introduces less than 1\% error on 82 of 85 datasets.}
    \label{fig:quantize_errs}
\end{center}
\end{figure}

To measure the error this introduced, we then inverted the linear transformation and computed the mean squared error between the original and the ``reconstructed'' data. The resulting error values for each dataset, normalized by the dataset's variance, are shown in Figure~\ref{fig:quantize_errs}. These normalized values can be understood as signal-to-noise ratio measurements, where the noise is the quantization error. As the figure illustrates, the quantization error is orders of magnitude smaller than the variance for nearly all datasets, and never worse than $10\times$ smaller, even for 8-bit quantization.

This of course does not indicate that all time series can be safely quantized. Two counterexamples of which we are aware are 1) timestamps where microsecond or nanosecond resolution matters, and 2) GPS coordinates, where small decimal places may correspond to many meters. However, the above results suggest that quantization is a suitable means of applying \minesp to floating-point data in many applications. This is bolstered by previous work showing that quantization even to a mere six bits \cite{epenthesis} rarely harms classification accuracy, and quantizing to two bits is enough to support many data mining tasks \cite{sax, hotSax, isax, saxvsm}.

\section{Conclusion} \label{sec:conclusion}

We introduce \mine, a compression algorithm for multivariate integer time series that achieves state-of-the-art compression ratios across a large number of publicly available datasets. It also attains speeds of up to 3GB/s in a single thread and predictable performance as a function of the number of variables being compressed. Moreover, it only needs to buffer eight samples at a time, enabling low latency for continuously arriving data. Finally, \minesp has extremely low memory requirements, making it feasible to run even on resource-constrained devices.

As part of evaluating \mine, we also conducted what is, to the best of our knowledge, the largest empirical investigation of time series compression that has been reported. To both ensure reproducibility of our work and facilitate future research in this area, we make available all of our experiments as a public benchmark.

In future work, we hope to characterize the relationship between compression and power savings, both for \minesp and for other methods. The savings are upper bounded by the compression ratio in the limit of data transmission consuming all power, but real-world systems have various overheads that cause significant deviation from this idealized model.

\section{Acknowledgements}

This material is based upon work supported by the National Science Foundation Graduate Research Fellowship under Grant No. 1122374. Any opinion, findings, and conclusions or recommendations expressed in this material are those of the authors(s) and do not necessarily reflect the views of the National Science Foundation.

\bibliographystyle{abbrv}
\bibliography{doc}

\balance

\end{document}